\documentclass[lettersize,journal,12pt]{IEEEtran}
\usepackage[T1]{fontenc}
\usepackage[utf8]{inputenc}
\usepackage{textcomp}
\usepackage{microtype}

\usepackage{amsmath,amssymb}
\usepackage{algorithmic}
\usepackage{algorithm}

\usepackage{array}

\usepackage{graphicx}
\usepackage[caption=false,font=normalsize,labelfont=sf,textfont=sf]{subfig}

\usepackage{booktabs}
\usepackage{multirow}
\usepackage{tabularx}

\usepackage[printonlyused]{acronym}

\acrodef{ai}[AI]{Artificial Intelligence}
\acrodef{xai}[XAI]{EXplainable Artificial Intelligence}
\acrodef{ml}[ML]{Machine Learning}
\acrodef{dl}[DL]{Deep Learning}
\acrodef{aml}[AML]{Adversarial Machine Learning}
\acrodef{nn}[NN]{Neural Networks}
\acrodef{dnn}[DNN]{Deep Neural Networks}
\acrodef{lstm}[LSTM]{Long-Short Term Memory}
\acrodef{gnn}[GNN]{Graph Neural Networks}
\acrodef{gan}[GAN]{Generative Adversarial Networks}
\acrodef{mae}[MAE]{Mean Absolute Error}
\acrodef{mse}[MSE]{Mean Square Error}

\acrodef{shap}[SHAP]{SHapely Additive exPlanations}
\acrodef{lime}[LIME]{Local Interpretable Model-agnostic Explanations}
\acrodef{xgb}[XGBoost]{Extreme Gradient Boosting}
\acrodef{lrp}[LRP]{LayeR-wise backPropagation}
\acrodef{fgsm}[FGSM]{Fast Gradient Sign Method}
\acrodef{bim}[BIM]{Basic Iterative Method}
\acrodef{dt}[DT]{Decision Tree}
\acrodefplural{DT}[DTs]{Decision Trees}

\acrodef{xrl}[XRL]{EXplainable Reinforcement Learning}
\acrodef{rl}[RL]{Reinforcement Learning}
\acrodef{drl}[DRL]{Deep Reinforcement Learning}
\acrodef{crl}[CRL]{Casual Reinforcement Learning}
\acrodef{dqn}[DQN]{Deep Q-Network}
\acrodef{ppo}[PPO]{Proximal Policy Optimization}
\acrodef{a3c}[A3C]{Asynchronous Advantage Actor-Critic}

\acrodef{ran}[RAN]{Radio Access Network}

\acrodef{mno}[MNO]{Mobile Network Operator}
\acrodef{bs}[BS]{Base Station}
\acrodefplural{BS}[BSs]{Base Stations}
\acrodef{gnb}[gNB]{next Generation NodeB}
\acrodefplural{gNB}[gNBs]{next Generation NodeBs}
\acrodef{enb}[eNB]{evolved NodeB}
\acrodefplural{eNB}[eNBs]{evolved NodeBs}
\acrodef{ue}[UE]{User Equipment}
\acrodef{ru}[RU]{Radio Unit}
\acrodef{du}[DU]{Distributed Unit}
\acrodef{cu}[CU]{Central Unit}

\acrodef{tbs}[TBS]{Tranport Block Size}
\acrodef{rrc}[RRC]{Radio Resource Control}
\acrodef{mcs}[MCS]{Modulation and Coding Scheme}
\acrodef{tti}[TTI]{Transmission Time Interval}
\acrodef{prb}[PRB]{Physical Resource Block}
\acrodefplural{PRB}[PRBs]{Physical Resource Blocks}
\acrodef{rnti}[RNTI]{Radio Network Temporary Identifier}
\acrodefplural{RNTI}[RNTIs]{Radio Network Temporary Identifiers}

\acrodef{embb}[eMBB]{enhanced Mobile Broadband}
\acrodef{mmtc}[mMTC]{enhanced Machine-type Communications}
\acrodef{urllc}[URLLC]{Ultra-Reliable and Low Latency Communications}
\acrodef{mcs}[MCS]{Modulation and Coding Scheme}

\acrodef{qos}[QoS]{Quality of Service}
\acrodef{kpi}[KPI]{Key Performance Indicator}
\acrodefplural{kpi}[KPIs]{Key Performance Indicators}
\acrodef{kpm}[KPM]{Key Performance Measurement}
\acrodefplural{kpm}[KPMs]{Key Performance Measurements}
\acrodef{sla}[SLA]{Service Level Agreement}
\acrodef{isp}[ISP]{Internet Service Providers}
\acrodef{uav}[UAV]{Unmanned Aerial Vehicle}
\acrodef{mdp}[MDP]{Markov Decision Process}

\acrodef{ric}[RIC]{RAN Intelligent Controller}
\acrodef{smo}[SMO]{Service Management and Orchestrator}

\acrodef{srn}[SRN]{Standard Radio Node}
\acrodef{rmr}[RMR]{RIC Message Router}

\acrodef{ht}[HT]{High-Throughput}
\acrodef{ll}[LL]{Low-Latency}

\acrodef{dwl}[DWL]{downlink}

\acrodef{qoe}[QoE]{Quality of Experience}
\acrodef{ott}[OTT]{Over-the-Top}

\acrodef{osc}[OSC]{O-RAN Software Community}
\acrodef{mgen}[MGEN]{Multi-Generator}

\acrodef{ci}[CI]{Continuous Integration}
\acrodef{cd}[CD]{Continuous Deployment}

\acrodef{zsm}[ZSM]{Zero-touch Service Management}

\usepackage{textcomp}
\usepackage{xurl}
\usepackage[most]{tcolorbox}
\usepackage{enumitem}

\usepackage{orcidlink}

\usepackage{fontawesome}

\usepackage{tikzpagenodes,etoolbox}
\usepackage[contents={}]{background}
\AddEverypageHook{%
	\ifnumequal{\thepage}{1}{%
		\tikz[remember picture,overlay]{%
			\node[draw,
			minimum width=\textwidth,
			text width=0.999\textwidth,
			font=\footnotesize,
			inner sep=0.5pt,
			] 
			at ([yshift=2ex]current page header area) 
			{%
				This is the author's accepted version of the article. The final version published by IEEE is 
				R. Soundrarajan, C. Fiandrino, M. Polese, Salvatore D'Oro, L. Bonati, and T. Melodia, ``On AI Verification in Open RAN,'' , IEEE COMMAG, 2025, pp. TBD, doi: TBD.
			};
			\node[draw,
			minimum width=\textwidth,
			text width=0.99\textwidth,
			font=\footnotesize,
			inner ysep=0.55pt,
			] 
			at (current page footer area) 
			{%
				©2025 IEEE. Personal use of this material is permitted. Permission from IEEE must be obtained for all other uses, in any current or future media, including reprinting/republishing this material for advertising or promotional purposes, creating new collective works, for resale or redistribution to servers or lists, or reuse of any copyrighted component of this work in other works.
			};
		}%
	}{}
}

\begin{document}
\bstctlcite{IEEEexample:BSTcontrol}

\title{On AI Verification in Open RAN}

\author{Rahul Soundrarajan$^{\orcidlink{0009-0006-6819-3031}}$, Claudio Fiandrino$^{\orcidlink{0000-0002-4323-4355}}$,~\IEEEmembership{Member,~IEEE,}
	Michele Polese$^{\orcidlink{0000-0002-9740-134X}}$,~\IEEEmembership{Member,~IEEE,}
	Salvatore D'Oro$^{\orcidlink{0000-0002-7690-0449}}$,~\IEEEmembership{Member,~IEEE,}\\
	Leonardo Bonati$^{\orcidlink{0000-0002-1511-1833}}$,~\IEEEmembership{Member,~IEEE,} and
	Tommaso Melodia$^{\orcidlink{0000-0002-2719-1789}}$,~\IEEEmembership{Fellow,~IEEE}
	\thanks{This work is partially supported by bRAIN project PID2021-128250NB-I00 funded by MCIN/ AEI /10.13039/501100011033/ and the European Union ERDF ``A way of making Europe''; C. Fiandrino is a Ramón y Cajal awardee (RYC2022-036375-I), funded by MCIU/AEI/10.13039/501100011033 and the ESF+. This work was also partially supported by the O-RAN ALLIANCE, and by the U.S.\ National Science Foundation under grant CNS-2112471.}
	\thanks{R.~Soundrarajan is with Tejas Networks, Bengaluru, India. E-mail: \{rahulsound\}\protect@mail.com.}%
	\thanks{C.~Fiandrino is with IMDEA Networks Institute, Madrid, Spain. E-mail: \{firstname.lastname\}\protect@imdea.org.}%
    \thanks{M.~Polese, S.~D'Oro, L.~Bonati, and T.~Melodia are with Northeastern University, Boston, MA, USA. E-mail: \{m.polese, l.bonati, s.doro, t.melodia\}\protect@northeastern.edu}%
    \vspace*{-5ex}%
 }

\markboth{IEEE Communications Magazine,~Vol.~xx, No.~xx, May~2025}%
{Soundrarajan~\MakeLowercase{\textit{et al.}}: On the AI Verifiability in the Open RAN}


\maketitle

\begin{abstract}
Open RAN introduces a flexible, cloud-based architecture for the \ac{ran}, enabling \ac{ai}/\ac{ml}-driven automation across heterogeneous, multi-vendor deployments. While \ac{xai} helps mitigate the opacity of AI models, explainability alone does not guarantee reliable network operations. In this article, we propose a lightweight verification approach based on interpretable models to validate the behavior of \ac{drl} agents for RAN slicing and scheduling in Open RAN. Specifically, we use \ac{dt}-based verifiers to perform near-real-time consistency checks at runtime, which would be otherwise unfeasible with computationally expensive state-of-the-art verifiers. We analyze the landscape of \ac{xai} and \ac{ai} verification, propose a scalable architectural integration, and demonstrate feasibility with a \ac{dt}-based slice-verifier. We also outline future challenges to ensure trustworthy \ac{ai} adoption in Open RAN.
\end{abstract}


\acresetall

\section{Introduction}
\label{sec:intro}

\IEEEPARstart{R}{ecent} years have witnessed growing interest in making the \ac{ran} more flexible and programmable. The Open \ac{ran} vision seeks to transform closed, hardware-bound architectures into virtualized, software-defined systems~\cite{comst-surveoran-wines}. The O-RAN ALLIANCE specifications operationalize this vision through an architecture composed of interoperable, multi-vendor components connected by standardized interfaces and the AI-RAN Alliance further promote the integration of \ac{ai}/\ac{ml} into the RAN ecosystem, fostering standardization and adoption of intelligent control mechanisms.

\ac{ai}/\ac{ml} technologies are emerging as key enablers in this ecosystem, powering closed-loop control mechanisms that optimize utility metrics (e.g., throughput) through trial-and-error. According to a 2024 SNS Telecom \& IT report~\cite{snstelecom-report}, investments in Open \ac{ran} automation are expected to grow by $125\,$\% by 2027, reaching nearly \$700 million. \ac{drl} techniques are increasingly used for \ac{ran} resource allocation, handover, load balancing~\cite{comst-drl-networking}, and misconfiguration handling~\cite{comnet-misconfig_oran}, thanks to their adaptability in distributed and dynamic environments. However, these models often operate as closed boxes, limiting interpretability and reducing operator trust. Researchers have adopted \ac{xai} techniques~\cite{conext-explora}, yet explainability without verifiability does not guarantee alignment with operational goals like \ac{sla} for latency and throughput, strict slice isolation, reliability and packet-loss bounds, fairness across users, and energy or cost budgets.

\ac{ai} verification complements \ac{xai} by checking whether AI models adhere to specified behaviors and perform reliably under expected conditions~\cite{comacm-trustai, acmtsem-verification} (e.g., in the absence of adversarial inputs). Through formal methods and testing, verification tackles safety, fairness, and robustness challenges that explanations alone cannot address. Unlike traditional approaches as model checking~\cite{urban2021review}, exhaustive state-space exploration~\cite{comacm-trustai}, or SMT-based verification~\cite{weng2019proven}, which are often computationally prohibitive, we introduce a lightweight, pragmatic verification strategy based on interpretable models like \acp{dt}. Our approach provides near-real-time feedback and is suitable for deployment alongside DRL-based xApps, such as those responsible for RAN slicing and scheduling.

In summary, this work addresses a key gap in Open \ac{ran} by integrating efficient verification into AI/ML-driven RAN control. Our contributions are threefold. First, we provide a critical overview of the roles of \ac{xai} and \ac{ai} verification, clarifying their interplay in the Open \ac{ran} context. Second, we introduce a system-level architectural mapping to embed verification into the O-RAN \ac{ai}/\ac{ml} lifecycle, identifying where telemetry and decisions can be introspected. Third, we present a use case where a slice-verifier based on \ac{dt} models verifies DRL-driven RAN slicing and scheduling, demonstrating its feasibility within RIC timing constraints. Finally, we outline future directions for scalable and trustworthy AI verification in Open \ac{ran} environments.

Overall, this paper aims to stimulate the research at the intersection of \ac{xai}, \ac{ai} verification and networking to continue improving \ac{ai} robustness for its adoption in Open \ac{ran}.

\section{{AI} Verification and XAI}
\label{sec:bck}

Verifying learning-based systems means checking whether a trained model satisfies specific behavioral or robustness criteria—such as resilience to adversarial inputs, tolerance to missing features, or generalization to unseen data. Traditional verification approaches often rely on formal tools like mixed-integer linear programming or logic-based solvers, but these lack scalability and generality across diverse scenarios~\cite{urban2021review,acmtsem-verification}. By contrast, explainability aims to make model behavior human understandable using methods such as decision rules, counterfactuals, or knowledge graphs~\cite{xai-comst}.

Verification and explainability are closely linked in learning-based systems, much like decision and optimization in computational complexity. Recent work shows that verification challenges (e.g., adversarial robustness) can inform explainability, and vice versa~\cite{aaai-xai_for_verification}. Unifying these concepts offers deeper insights into AI decision-making and helps prevent errors before deployment.

\subsection{Applications and Benefits of XAI}
\label{subsec:xai-openran}

\noindent\textbf{XAI in Brief.} Explainability is key to building trust in AI systems. XAI techniques fall into three main categories: post-hoc, counterfactual, and model-inherent. Post-hoc methods such as \ac{lime}, \ac{shap}, \ac{lrp}, and GRAD-CAM analyze models after training and can be applied to any closed-box model. Surrogate models like \acp{dt} also belong to this group. Counterfactual approaches explain decisions by showing how small input changes could alter predictions, offering insights into decision boundaries. Model-inherent techniques use self-interpretable models (e.g., \acp{dt}) or design models with interpretability embedded in the training process. Coupling explainability with robustness is key for telecom resource management and explainable models should be designed to operate within closed-loop automation frameworks such as \ac{zsm}~\cite{xaicommag}.

\noindent\textbf{Challenges of XAI in the Open RAN.} The Open \ac{ran} framework lays the path forward for closed-loop control, with the \ac{ran} exposing context and telemetry to controllers and third-party \ac{ai}/\ac{ml} applications that dynamically tune and optimize the network. In this context, the development of efficient control mechanisms remains an open research area. As the industry transitions toward more intelligent control, there is a pressing need for robust, reliable, and deployable ML solutions specifically tailored to wireless networks.  However, the existing solutions often rely on a combination of heuristics and \ac{ai}/\ac{ml} techniques. These include optimizing training methods for dynamic online systems, developing \ac{drl} solutions that generalize well across diverse deployments, and exploring the potential of generative AI in the Open \ac{ran} context~\cite{comst-surveoran-wines}. 

Unfortunately, achieving explainability in systems like cellular networks with dynamic control and complex input/output relationships is complex:

\begin{enumerate}[label=$\bullet$,
wide=0\parindent,
listparindent=0pt, 
align=left]
\item The \ac{ran} is a non-stationary and dynamic environment with varying operational conditions such as channel propagation characteristics and traffic profiles. As these conditions are very hard to model, the generalizability of model-free \ac{ai}/\ac{ml} techniques is key for effective solutions. At the same time, this makes the generation of explanations that are informative and consistent with the current dynamics of the system challenging.
\item For the explanations to be effective, the computational complexity of the \ac{xai} techniques must be kept under control. In the \ac{ran}, operations occur at timescales that stem from few milliseconds to several seconds. However, techniques like \ac{shap} could take up to several hours to produce global explanations~\cite{conext-explora}.
\end{enumerate}

\noindent\textbf{Benefits of XAI in the Open RAN.}	In the context of Open \ac{ran}, specific benefits of \ac{xai} include:

\begin{enumerate}[label=$\bullet$,
wide=0\parindent,
listparindent=0pt, 
align=left]
	\item \textit{Troubleshooting and monitoring} of the \ac{ai}/\ac{ml} models to prevent misconfigurations and help with conflict mitigation. The Open \ac{ran} ecosystem is heterogeneous with hardware and software from diverse manufacturers and different radio access technologies~\cite{comnet-misconfig_oran}. Therefore, automating network operations via \ac{ai}/\ac{ml} could lead to misconfigurations that are hard to troubleshoot and resolve without deep knowledge about the \ac{ai}/\ac{ml} decisions.
	\item \textit{Intent-based networking} aims at enabling simplified and agile network management where complex configurations are translated into high-level intents such as guarantee  \ac{urllc} 95-th-percentile latency $\leq x$ ms. In this context, \ac{xai} can shed light whether the models responsible for network configuration are consistently enforcing the given intents~\cite{conext-explora}.	
\end{enumerate}

\subsection{The Current Landscape of AI Verification}
\label{subsec:bck-ai-verification}

Formal methods have been successfully applied to verify classical software systems and only in recent years researchers have started to investigate their application to \ac{ai}/\ac{ml}~\cite{urban2021review,comacm-trustai}. Vis-a-vis with the classical software systems, \ac{ai}/\ac{ml} systems pose specific challenges. Beyond being inherently complex and difficult to understand, \ac{ai}/\ac{ml} solutions for cellular networks are also significantly affected by the unpredictable nature of wireless channels and struggle to perform well on data different from what they were trained on.

The landscape of \ac{ai} verification includes \textit{probabilistic} and \textit{non-probabilistic} methods~\cite{urban2021review,weng2019proven}. The latter were originally conceived for software with deterministic input-output nature and strive to prove that the outputs of an \ac{ai} model satisfy given criteria for all inputs in a given input space. Such techniques are Satisfiability Modulo Theories (SMT), Mixed Integer Linear Programming (MILP), reachability analysis. Although these techniques are very effective, they scale poorly~\cite{urban2021review}. By contrast, probabilistic techniques are tailored to AI systems and aim either to find upper bounds on the probability of respecting/violating a given criteria or to estimate the confidence of the output of the model given the statistical properties of the input. These techniques are more scalable at the expense of soundness in executing verification operations~\cite{weng2019proven,acmtsem-verification}.

The class of \ac{ai}/\ac{ml} models that are of interest for the Open \ac{ran} encompasses deep learning algorithms for regression and classification tasks, and \ac{drl}~\cite{comst-surveoran-wines}. Verifying regressors and classifiers is an important task, but it is often of a different scope than verifying DRL agents deployed for RAN control.  The latter directly impact network operations while the former classes of models require additional processing before acting upon network mechanisms. Verifying the operation of \ac{drl} involves the assessment of the degree of safety of the actions that the agent takes, i.e., if there is no harm on the state and whether there exist variations in the effect of one action enforced at different times. The verification process for \ac{drl} agents is complex because the output of one decision becomes the input for the next, creating a cumulative effect which adds exponential complexity to verification tools as the number of time steps increases~\cite{acmsurv-drl_verification}.

\section{{AI} Verification in the Open RAN}
\label{sec:arch}

\begin{figure*}[t]
    \centering%
    \includegraphics[width=.65\linewidth,keepaspectratio]{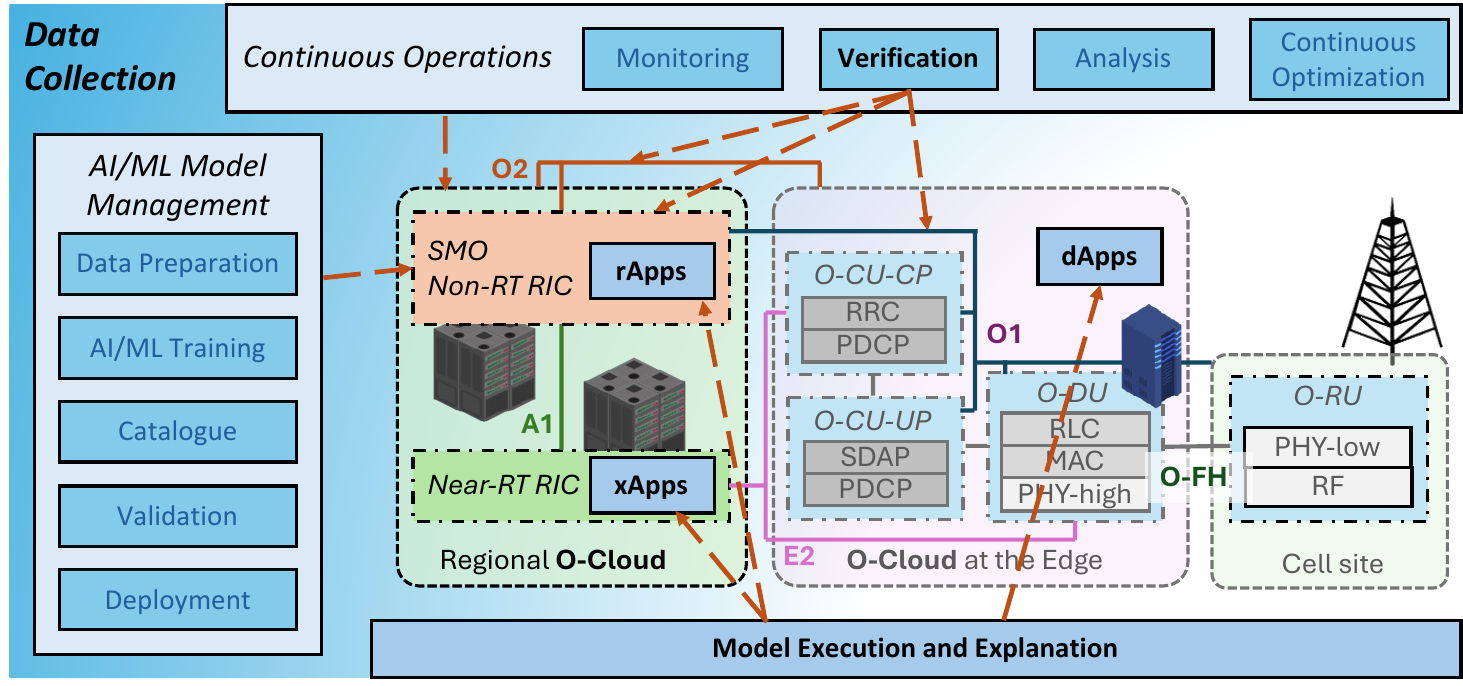}%
    \vspace*{-1ex}%
    \caption{O-RAN architecture and \ac{ai}/\ac{ml} workflows, focusing on model execution, explainability, and verification.} 
    \label{fig:o-ran-architecture}%
    \vspace*{-2ex}%
\end{figure*}

Verifying \ac{ai} behavior in Open RAN is essential to ensure that autonomous decisions made by \ac{ai}/\ac{ml} models align with operational goals like \ac{sla} assurance. Within the O-RAN architecture, verification must occur continuously and transparently, without disrupting ongoing network operations. This section outlines how \ac{ai} verification integrates into the existing O-RAN workflow, and where verification modules can be deployed.

\ac{ai} and \ac{ml} play a central role in the O-RAN architecture, which has been designed to support the \ac{ai}/\ac{ml} lifecycle from data collection and training to deployment and inference or control based on live data from the network, as shown in Fig.~\ref{fig:o-ran-architecture}. This is facilitated by a set of open interfaces, which expose data, telemetry, and control from \ac{ran} nodes (primarily, the \ac{cu}, \ac{du}, and \ac{ru}) to the \acp{ric} and the \ac{smo}. Specifically, the \acp{ric} onboard custom control logic and operate at near-real-time and non-real-time time scales---between 10\,ms and 1\,s (with custom applications called xApps), or above 1\,s (with rApps), respectively---and enforce configurations for the radio resources or deploy policies for system management and optimization. Real-time control (below 10\,ms) is provided by dApps, in an extension of the O-RAN architecture. These components are deployed on a heterogeneous set of compute resources, known as the O-Cloud.

The O-RAN architecture supports several functionalities associated with the lifecycle of \ac{ai}/\ac{ml} applications~\cite{oran-wg2-ml}, which can be deployed across different endpoints of the architecture. The \ac{smo} usually supports data-lake functionalities, collecting telemetry and logs from the thousands of devices and network functions under its purview. It can also host data preparation and model training pipelines, together with a model catalog to deploy trained solutions on the infrastructure (left part of Fig.~\ref{fig:o-ran-architecture}). 
The Non-RT \ac{ric}, often co-located with the \ac{smo}, can share some of these responsibilities, together with continuous operations in support of the monitoring and deployment of the models on the infrastructure. Finally, both \acp{ric} and the \acp{cu} and \acp{du} host model inference, as rApps, xApps, or dApps in the real-time extension of the O-RAN architecture. 

When it comes to explainability and verification, rApps, xApps, or dApps often share the need to access to a similar set of telemetry (e.g., \acp{kpi} from different elements of the RAN, actions and decisions taken by the deployed \ac{ai}/\ac{ml} models) and report to a similar set of stakeholders. These include telco personnel, as well as \ac{ci} and \ac{cd} pipelines that manage the models and their deployment. The latter is a critical element in the architecture, as the autonomous and automated nature of O-RAN systems calls for a closed-loop solution for the management of \ac{ai}/\ac{ml} models. If verification fails, the model is not apt for controlling or optimizing the network and must be updated, undeployed, or, in more severe cases, replaced entirely or flagged for human operator intervention. The verification results may change dynamically, as the actual network environment and the one known to the model may drift.

\begin{figure*}
\centering%
\includegraphics[width=.85\textwidth,keepaspectratio]{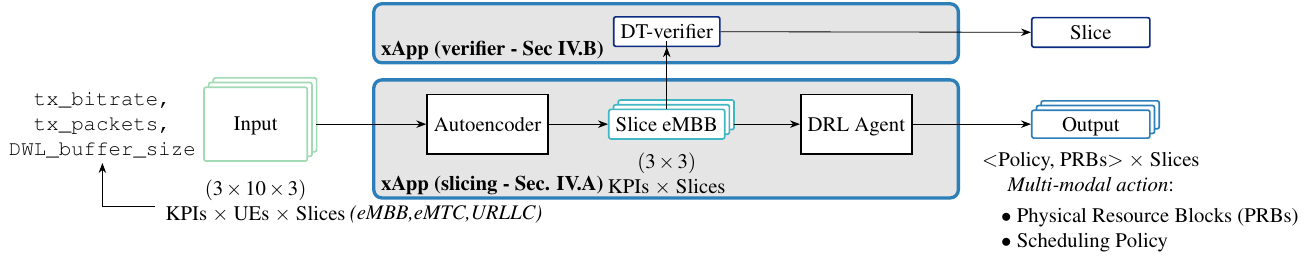}%
\vspace*{-3ex}%
\caption{Operation of the \ac{drl} agent xApp and the DT-based verifier xApp}%
\label{fig:drl-arch}%
\vspace*{-3ex}%
\end{figure*}

There are, however, differences in where verification and explainability algorithms are deployed. Explainability solutions, such as~\cite{conext-explora}, are often coupled with the models, e.g., as extensions in xApps, rApps or components in \acp{ric}, to directly interact with the same \acp{kpi} data. Verification techniques, instead, track performance trends over time across components, models, and infrastructure, and thus naturally fit into operations and O-Cloud components. Depending on their function, verifiers can be integrated at multiple layers. For instance, the use case in Sec.~\ref{subsec:verification} relies on analyzing large-scale datasets, multiple model instances, and telemetry. For near-real-time verification of inference outputs, lightweight verifiers like our \ac{dt}-based solution can be co-located with xApps in the Near-RT RIC, enabling monitoring and feedback at sub-second timescales. Other techniques without such tight timing requirements may access data lakes or infrastructure \acp{kpi} to capture overall \ac{ran} behavior rather than endpoint decisions. In addition, verification pipelines can also be validated in pre-deployment environments, e.g., digital twins or testbeds.

\section{The Verification Use Case}
\label{sec:use-case}

\subsection{The DRL Agents}
\label{subsec:agents-use-case}

We showcase the use of \ac{xai} for the verification of the behavior of a relevant \ac{ai}/\ac{ml} application for O-RAN systems. Specifically, we consider the case of joint control of \ac{ran} slicing and scheduling policies for a set of 5G slices: \textit{(i)} \ac{embb}; \textit{(ii)} \ac{mmtc}; and \textit{(iii)} \ac{urllc}. For each slice, the \ac{drl} agent~\cite{tmc-coloran-wines} selects a \ac{ran} slicing policy (i.e., the number of \acp{prb} reserved to the slice), and the optimal scheduling policy to serve the \acp{ue} of the slice among Round Robin (RR), Waterfilling (WF), and Proportional Fair (PF). Fig.~\ref{fig:drl-arch} shows the agent that is implemented as an xApp and provides the appropriate configuration to the \ac{gnb} each $250$\,ms.

Specifically, the bottom of Fig.~\ref{fig:drl-arch} depicts the architecture of the \ac{drl} agent that takes actions to maximize a target reward by monitoring a set of \acp{kpi} received from the \ac{gnb} via the E2 interface and processed by an autoencoder. The \acp{kpi} are transmission bitrate in Mbps, number of transmitted packets, and size of the \ac{dwl} buffer in bytes. These metrics are collected as individual measurements for each user of every slice.

In this work, we leverage a public dataset~\cite{conext-explora} that analyzes two different \ac{drl} agents: the \ac{embb}-oriented agent strives to maximize the transmission bitrate, while the \ac{urllc}-oriented agent minimizes \ac{dwl} buffer occupancy, which acts as a proxy for latency. The dataset contains per-user information on the above-mentioned \acp{kpi}, among others, as well as logs of the agents' operation (observed state, action and reward) in an urban scenario emulated with the Colosseum Open \ac{ran} digital twin. We refer the reader to~\cite{conext-explora} for the complete details of the dataset. Since the dataset provides a low number of samples for the \ac{embb}-oriented agent, we augmented the samples to be $10\,000$ via Conditional Tabular GAN (CTGAN), a generative model tailored for tabular data that preserve the joint KPI–slice distributions.
\vspace*{-2ex}

\subsection{Verifying Agent Behavior with Decision Trees}
\label{subsec:verification}

For our verification use-case, we address the following question:
\begin{tcolorbox}[boxrule=0pt,frame hidden,sharp corners,enhanced,borderline west={3pt}{0pt}{red!80!black}]
\textit{Is it possible to infer which is the slice a user belongs to and whether the agent decision is adequate for that slice by only observing the input \acp{kpi} (to the model) of a user?}
\end{tcolorbox}
\vspace*{-2pt}
The question may appear obvious as the mobile network operator has full knowledge about the user-to-slice mapping. However, this simple question makes it possible to verify whether the agent operation is consistent with the knowledge it acquired during training. Since the agent was trained to perform joint \ac{ran} slicing and scheduling control over the three slices simultaneously, each slice has a unique fingerprint in terms of resource assignment. The capability of verifying the agent's behavior is instrumental to troubleshoot improper or conflicting configurations generated by different xApps with diverse reward functions and target \acp{kpi}.

For verification, we propose to use \acp{dt}, a popular method commonly used for classification tasks. \acp{dt} classify a population into branch-like segments that construct an inverted tree with a root node, internal nodes, and leaf nodes. Being a non-parametric algorithm, \acp{dt} efficiently manage large datasets without imposing a parametric structure, handle well both categorical and numerical data, and are robust to outliers and missing data which is a common case in \ac{ran} telemetry. \acp{dt} align well with RAN constraints: they are fast to train and infer, and naturally interpretable. Unlike attribution methods such as SHAP or LRP, which often incur significant computational overhead, \acp{dt} enable slice verification at near-real-time timescales, making them deployable alongside xApps in the Near-RT RIC. Compared to neural-symbolic approaches or probabilistic verification, DTs offer a pragmatic balance between interpretability and operational feasibility at stringent timescales. 

\begin{figure*}
	\centering%
	\includegraphics[width=.87\textwidth,keepaspectratio]{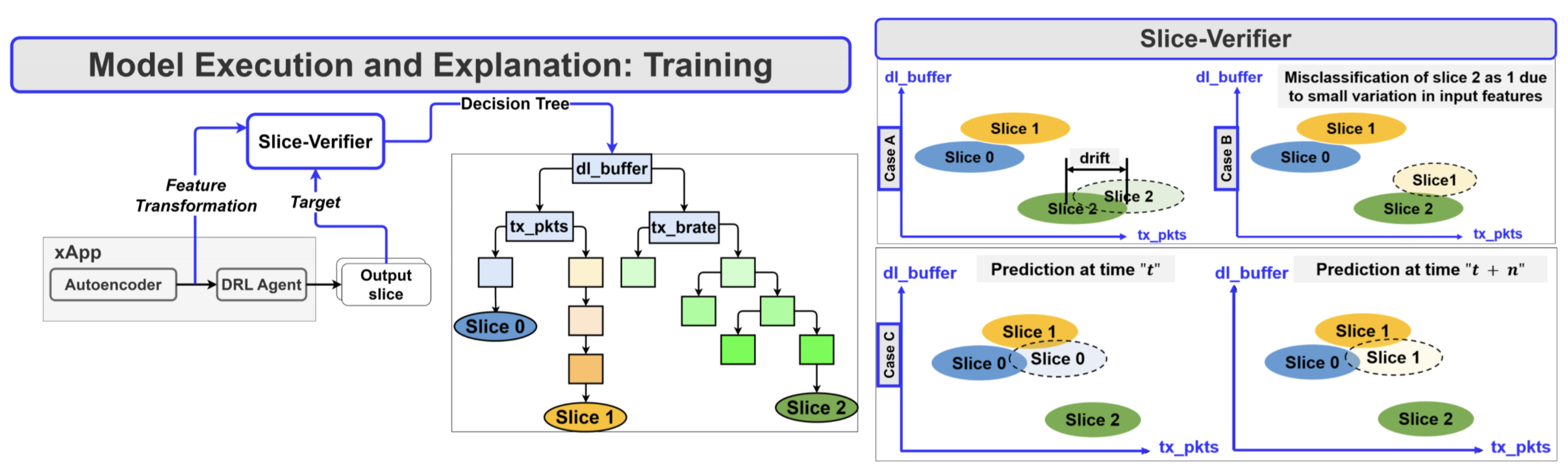}%
	\vspace*{-2ex}%
	\caption{The slice-verifier: training and verification capabilities. Case A shows a ``drift'' in the feature space for a specific slice, case B highlights a misclassification, and case C illustrates conflicting predictions for data points in a similar feature space at different times.
    }%
	\label{fig:obj-verifier}%
    \vspace*{-3ex}%
\end{figure*}

The top of Fig.~\ref{fig:drl-arch} and Fig.~\ref{fig:obj-verifier} illustrate how the slice-verifier is integrated within the normal operation of the \ac{drl}-based xApp, and showcases its capabilities in a few examples. Specifically, the slice-verifier can identify and signal \textit{a)} drifts in feature space for a specific slice, \textit{b)} misclassifications based on trained data can identify bias or error in learning, and \textit{c)} conflicting predictions for similar feature space that might be caused due to approximations or coarse feature transformations. We train the slice-verifier with XGBoost using as input features those of the agent and as output classes the labels of the slices. We tune XGBoost and \ac{dt} hyperparameters to balance expressiveness (by capturing non-linear relations in KPIs and ensuring all features contribute per splits) and lightweight inference (by restricting the maximum depth and minimum samples per split).

To demonstrate that the verification could be effective and lightweight, we restrain our support to train the slice-verifier to a small subset (less than 0.1\%) of samples that were used to train the agent. We test the accuracy using using standard classification metrics like ``Precision'', ``Recall'' and ``F1-score''. Tables~\ref{tab:acc-dt-embb} and Table~\ref{tab:acc-dt-urllc} show the accuracy of the \ac{dt} slice-verifier for the \ac{embb}- and \ac{urllc}-oriented agents in all slices. We observe that the agents' operation is robust because the slice-verifier achieves high accuracy in all the metrics, meaning that it is capable to correctly identify samples and avoid false positives. Further, across all the experiments that were carried out with a varying number of users, the variance of the input \ac{kpi} is low. The accuracy is higher for the \ac{urllc}-oriented agent as its reward function yields \ac{kpi} distributions with clear patterns (low buffer). In contrast, the \ac{embb}-oriented agent faces more trade-offs to optimize throughput, yielding \ac{kpi} distributions with some overlap with the \ac{mmtc} slice, which makes the verification harder and explains the lower accuracy values.

\begin{table}[t]
\centering%
\caption{Metrics of slice-verifier for \ac{embb}-oriented agent}%
\label{tab:acc-dt-embb}%
\vspace*{-2ex}%
\resizebox{\columnwidth}{!}{%
\begin{tabular}{lrrrr}
\toprule
\textsc{Slice} & \textsc{Precision} & \textsc{Recall} & \textsc{F1-Score} & \textsc{Support} \\ 
\midrule
\ac{embb} & 0.81 & 0.81 & 0.81 & 3297 \\
\ac{mmtc} & 0.82 & 0.82 & 0.82 & 3479 \\
\ac{urllc} & 0.81 & 0.81 & 0.81 & 3224 \\ \hline
\textit{Accuracy} & \multicolumn{2}{r}{0.82} & & 10000 \\ \hline
\textit{Macro Avg.} & 0.82 & 0.82 & 0.82 & 10000 \\
\textit{Weighted Avg.} & 0.82 & 0.82 & 0.82 & 10000 \\
\bottomrule
\end{tabular}%
}%
\vspace*{-2ex}%
\end{table}

\begin{table}[t]
\centering%
\caption{Metrics of slice-verifier for \ac{urllc}-oriented agent}%
\label{tab:acc-dt-urllc}%
\vspace*{-2ex}%
\resizebox{\columnwidth}{!}{%
\begin{tabular}{lrrrr}
\toprule
\textsc{Slice} & \textsc{Precision} & \textsc{Recall} & \textsc{F1-Score} & \textsc{Support} \\
\midrule
\ac{embb} & 0.99 & 0.98 & 0.99 & 3392 \\
\ac{mmtc} & 0.86 & 0.87 & 0.87 & 3280 \\
\ac{urllc} & 0.87 & 0.87 & 0.87 & 3328 \\ \hline
\textit{Accuracy} & \multicolumn{2}{r}{0.91} & & 10000 \\ \hline
\textit{Macro Avg.} & 0.91 & 0.91 & 0.91 & 10000 \\
\textit{Weighted Avg.} & 0.91 & 0.91 & 0.91 & 10000 \\
\bottomrule
\end{tabular}%
}%
\vspace*{-2ex}%
\end{table}

While DTs offer lightweight interpretability, they also come with limitations. They may oversimplify complex, high-dimensional feature relationships that advanced models, instead, capture. Simplicity is a strength in our case, but may not generalize to all AI/ML applications within Open RAN. Moreover, DT-based verification focuses on consistency in behavior, rather than formal guarantees. Future work should investigate hybrid methods, e.g., combining DTs with formal verification or counterfactual reasoning, to enhance trustworthiness.

\section{Open Challenges and Future Directions}
\label{sec:discussion}

As the need for network automation within the Open \ac{ran} grows, so does the importance of ensuring the trustworthiness of \ac{ai}/\ac{ml} operations. This section explores fundamental challenges ahead for AI verification and XAI, including tailoring the verification to Open \ac{ran} tasks, scalability of the proposed solutions, and prototyping and testing.

\paragraph{Open \ac{ran}-Oriented Verification}
\label{subsec:openran-verification}

The current approach for AI verification, which focuses on individual models in isolation, is insufficient for the RAN ecosystem's numerous interacting components~\cite{acmtsem-verification,urban2021review,weng2019proven}. Therefore, it is necessary to develop verification techniques that extend beyond a single \ac{ai}/\ac{ml} model to provide system-level guarantees. Standardizing the integration of explainability and robustness checks into frameworks like \ac{zsm} is an important research direction to enable trustworthy automated resource management~\cite{xaicommag}.

\paragraph{Tradeoff Scalability-Soundness}
\label{subsec:comput-complexity}

The computation time for highly sound verification techniques is very high, making them impractical for real-time operations. A promising way to overcome this challenge is to define system-specific boundaries to reduce the solving space and use explainability to understand the limits of the model's operation.

\paragraph{Prototyping and Testing}
\label{sec:proto-testing}
Prototyping and testing verification and XAI solutions requires a phased approach, involving multiple resources from testbeds used as sandbox for development, to production network environments that can offer access to data at scale. This is necessary to test the scalability of proposed AI verification techniques, and of the control and optimization architectures enabling them.

\section{Concluding Remarks}
\label{sec:concl}

This paper analyzed \ac{xai} and \ac{ai} verification techniques and outlined their applicability to Open RAN. The unique properties of the RAN ecosystem call for tailored verification approaches that can operate across multi-vendor systems and at timescales compatible with RAN control. Through our use case, we showed that DRL decisions for slicing and scheduling can be effectively verified using lightweight \ac{dt}-based models, ensuring consistency with training knowledge. While \ac{xai} and \ac{ai} verification in Open RAN are still in their early stages, our study provides a foundation for advancing trustworthy \ac{ai}/\ac{ml} adoption in this domain.

\bibliographystyle{IEEEtran}
\bibliography{biblio.bib}

 \section*{Biographies}

\vspace*{-4ex}

\begin{IEEEbiographynophoto}{Rahul Soundrarajan} is a Principal Engineer at Tejas Networks. Prior to this, he was an entrepreneur \& independent  consultant. He has designed, developed \& evaluated ML Algorithms for Near-RT RIC, Non-RT RIC and network optimization use cases for which he holds several patents. Contributions to this work was done by Rahul as an independent research consultant prior to joining Tejas Networks. 
\end{IEEEbiographynophoto}
\vspace*{-7ex}

\begin{IEEEbiographynophoto}{Claudio Fiandrino} is a Research Assistant Professor at IMDEA Networks Institute, Spain, where he leads the Laboratory for Resilient AI Networking. His expertise and interests lie at the interface of explainable and robust \ac{ai}/\ac{ml} and mobile networks.
\end{IEEEbiographynophoto}
\vspace*{-7ex}

\begin{IEEEbiographynophoto}{Michele Polese} is a Research Assistant Professor at the Institute for the Wireless Internet of Things at Northeastern University, Boston. He received his Ph.D. in Information Engineering at the University of Padova in 2020. He then joined Northeastern University as a research scientist. His research interests are in the analysis and development of protocols and architectures for future generations of cellular networks.
\end{IEEEbiographynophoto}
\vspace*{-7ex}

\begin{IEEEbiographynophoto}{Salvatore D'Oro} is a Research Associate Professor with Northeastern University. He received his Ph.D.\ from the University of Catania in 2015. His research focuses on optimization and learning for NextG systems and the Open RAN. 
\end{IEEEbiographynophoto}
\vspace*{-7ex}

\begin{IEEEbiographynophoto}{Leonardo Bonati} is an Associate Research Scientist at the Institute for the Wireless Internet of Things, Northeastern University, Boston. He received a Ph.D.\ degree in Computer Engineering from Northeastern University in 2022. His research focuses on softwarized approaches for the Open RAN of next generation of cellular networks.
\end{IEEEbiographynophoto}
\vspace*{-7ex}

\begin{IEEEbiographynophoto}{Tommaso Melodia}
received a Ph.D.\ in Electrical and Computer Engineering from the Georgia Institute of Technology in 2007. He is the William Lincoln Smith Professor at Northeastern University, the Director of the Institute for the Wireless Internet of Things, and the Director of Research for the PAWR Project Office. His research focuses on wireless networked systems.
\end{IEEEbiographynophoto}

\end{document}